\documentclass[reprint, 
 amsmath,amssymb,
 aps,
]{revtex4-1}

\usepackage[a4paper, margin=2cm]{geometry}
\usepackage{graphicx}
\usepackage{dcolumn}
\usepackage{bm}
\usepackage[mathlines]{lineno}

\usepackage{upgreek}

\begin{document}


\title{Controlling Anomalous Diffusion in Lipid Membranes}

\author{H. L. E. Coker*}
\author{M. R. Cheetham*}
\affiliation{*Contributed equally to this work.}
\affiliation{Department of Chemistry, Britannia House, King's College London, London SE1 1DB}
\affiliation{Chemistry Research Laboratory, 12 Mansfield Road, University of Oxford, Oxford OX1 3TA}
\author{D. R. Kattnig}
\affiliation{Living Systems Institute \& Department of Physics, University of Exeter, Stocker Road, Exeter EX4 4QD}
\author{Y. J. Wang}
\author{S. Garcia-Manyes}
\affiliation{Department of Physics, Strand Building, King's College London, London WC2R 2LS}
\author{M. I. Wallace}
\affiliation{Department of Chemistry, Britannia House, King's College London, London SE1 1DB}
\date{\today}

\begin{abstract}
Diffusion in cell membranes is not just simple two-dimensional Brownian motion, but typically depends on the timescale of the observation. The physical origins of this anomalous sub-diffusion are unresolved, and model systems capable of quantitative and reproducible control of membrane diffusion have been recognized as a key experimental bottleneck. Here we control anomalous diffusion using supported lipids bilayers containing lipids derivatized with polyethylene glycol (PEG) headgroups. Bilayers with specific excluded area fractions are formed by control of PEG-lipid mole fraction. These bilayers exhibit a switch in diffusive behavior, becoming anomalous as bilayer continuity is disrupted. Diffusion in these bilayers is well-described by a power-law dependence of the mean square displacement with observation time. The parameters describing this diffusion can be tailored by simply controlling the mole fraction of PEG-lipid, producing bilayers that exhibit anomalous behavior similar to biological membranes.  
\end{abstract}

\maketitle

\section{Introduction}

Diffusion is an essential transport mechanism in membrane biology, vital for a wide range of biological function including protein organization \cite{Sheets1997}, signalling \cite{Choquet2003,Kholodenko2006} and cell survival \cite{Cheema2012}. Interestingly, such living systems do not in general display the Brownian motion predicted by a simple random walk model, and instead exhibit `anomalous' diffusion \cite{Saxton1994} where the diffusivity is dependent on the timescale of observation. This phenomenon has been reported both for three-dimensional diffusion in the cell cytosol \cite{Regner2013} and two-dimensional diffusion in the plasma membrane \cite{Hofling2013, Fujiwara2016, Golan2017}. Here we focus on membrane diffusion.

Why and how anomalous diffusion exists in the plasma membrane has been the subject of considerable investigation (reviewed in \cite{Saxton2012}). The common underlying mechanism is thought to be the crowded environment found in the cell membrane \cite{Kusumi2005}, and the presence of slower-moving obstacles \cite{Saxton1987,Berry2014}, pinning sites, and compartmentalization \cite{Fujiwara2002,Murase2004,Kusumi2005} have all been suggested as potential contributors to anomaleity in membrane diffusion. Confinement in cellular membranes is observed on the order of tens to hundreds of nanometers, with anomalous diffusion reported in a large number of cell types \cite{Fujiwara2002,Murase2004}. Overall this work has led to the adoption of a compartmentalized `picket fence' model of the cell membrane as a proposed improvement to the `fluid-mosaic' model \cite{Kusumi2005}.

Artificial lipid bilayers have played a key role in improving our understanding of anomalous diffusion \cite{Schutz1997,Ratto2003,Horton2010,Spillane2014, Wu2016,Rose2015}, where both phase separation \cite{Ratto2003} and protein binding \cite{Horton2010} in supported lipid bilayers (SLBs) have been used to generate anomalous diffusion.  Simulations have also been vital in advancing our understanding, with much pioneering \cite{Saxton1989,Saxton1994,Saxton2001} and recent \cite{Stachura2014,Mardoukhi2015,Koldso2016,Jeon2016,Bakalis2015,Javanainen2013} work in this area. In particular, simulations have helped elucidate the role of mobile and immobile obstacles in causing anomalous behavior \cite{Saxton1987,Berry2014}. Relevant to our work, simulations have also been used to better interpret single particle tracking data \cite{Kepten2015} and provide methods to discriminate between classes of anomalous diffusion \cite{Metzler2014}. 

Despite these advances, the specific molecular mechanisms that give rise to anomalous diffusion \emph{in vivo} remain elusive. This is most clearly highlighted by Saxton et al., who published a call for `a positive control for anomalous diffusion' as a solution to this problem \cite{Saxton2012}. This positive control would be a simple and reproducible experimental model exhibiting `readily tuneable' anomalous diffusion spanning several orders of magnitude in timescale. Here we seek to address this call by engineering a simple experimental model in which it is possible to select the anomalous behavior. We take advantage of previous work on the disruption of SLB formation by PEG-DPPE \cite{Kaufmann2009} to control nanoscale obstacle formation in a bilayer (Fig. 1A). By varying the PEG-DPPE composition in a bilayer, we expect that an increased fraction of polymer in the brush regime will result in the formation of specific defects in the bilayer, similar to interfacial or grain boundary defects caused by phases separating mixtures \cite{Keller2005}. Similar defects have also been reported using Atomic Force Microscopy (AFM) of incomplete SLB formation from small unilamellar vesicles (SUVs) \cite{Richter2005} as well as in SLBs formed in the presence of membrane active peptides \cite{Oliynyk2007}.

The complex nanometer-scale confinement reported in cell membranes gives rise to anomalous behavior that spans from microseconds to seconds. Thus to properly characterize anomalous diffusion, it is important to apply techniques capable of studying these timescales. Here we exploit a combination of single-molecule Total Internal Reflection Fluorescence (smTIRF) \cite{Axelrod:1984aa} and Interferometric Scattering (iSCAT) microscopies \cite{Lindfors2004, Ortega-Arroyo2012a} (Figs. 1B and C, Supplementary Methods) to characterize diffusion using single-particle tracking that spans over four orders of magnitude in time.

\begin{figure}
\includegraphics[width=0.9\linewidth]{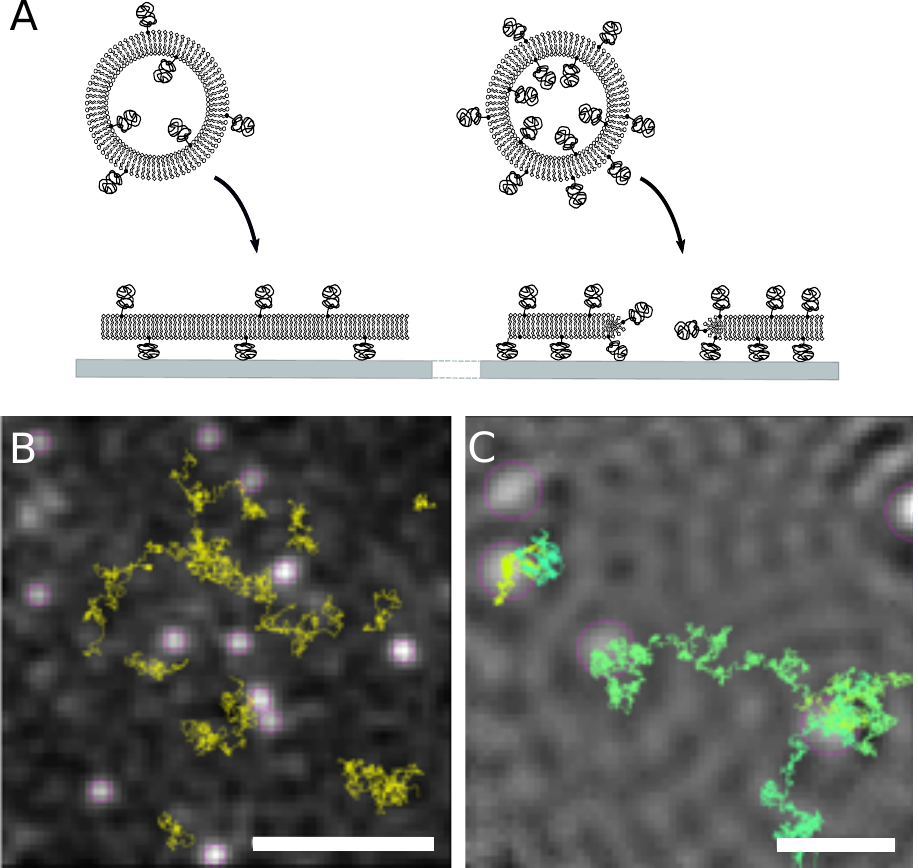}
\caption{\textbf{PEG bilayer model}. (\textbf{A}) Schematic of supported lipid bilayers. As the mole fraction of PEGylated lipids increases (left to right), defects form in the bilayer that that act as obstacles, generating anomalous diffusion. Representative single-particle tracking of smTIRF (\textbf{B}) and iSCAT (\textbf{C}) images (scale 10 $\upmu$m and 1 $\upmu$m, respectively).}	
\end{figure}

\section{Theory}
Anomalous diffusion describes random molecular motion that does not display a linear scaling of the second moment with time. The most common model for anomalous diffusion is to allow the second moment to scale as a power of time \cite{Havlin1987,Saxton1994},
\begin{equation}
\label{eqn2}
	\displaystyle
	\left<\Delta r^2\right> = 4 \Gamma \Delta t^\alpha, 
\end{equation}
where $\alpha$ is the anomalous exponent and $D$ is replaced by $\Gamma$, the anomalous transport coefficient. Anomalous sub- and super-diffusion are defined by  $\alpha < 1$ and  $\alpha > 1$. Given the form of equation \ref{eqn2}, $\alpha$ can be determined from the gradient of a logarithmic plot of $\left<\Delta r^2\right>/4\Delta t$ \emph{vs.} $\Delta t$.

The transport coefficient $\Gamma$ is somewhat more difficult to interpret as it has dimension of [L]$^2$/[T]$^\alpha$, thus its dimensions are changing for different degrees of anomalous behavior. This apparent problem can be overcome by de-dimensionalizing the observation time \cite{Saxton1994} using a `jump time', $\tau$:
\begin{equation}
\label{eqn3}
	\displaystyle
	\left<\Delta r^2\right> = 4 D \Delta t \left(\frac{\Delta t}{\tau}\right)^{\alpha-1}.
\end{equation}\\
$\tau$ can be interpreted in terms of a length scale ($\lambda$) associated with the anomalous behavior in 2D ($\lambda=\sqrt{4D\tau}$).

\begin{figure*}
	\includegraphics[width=0.9\linewidth]{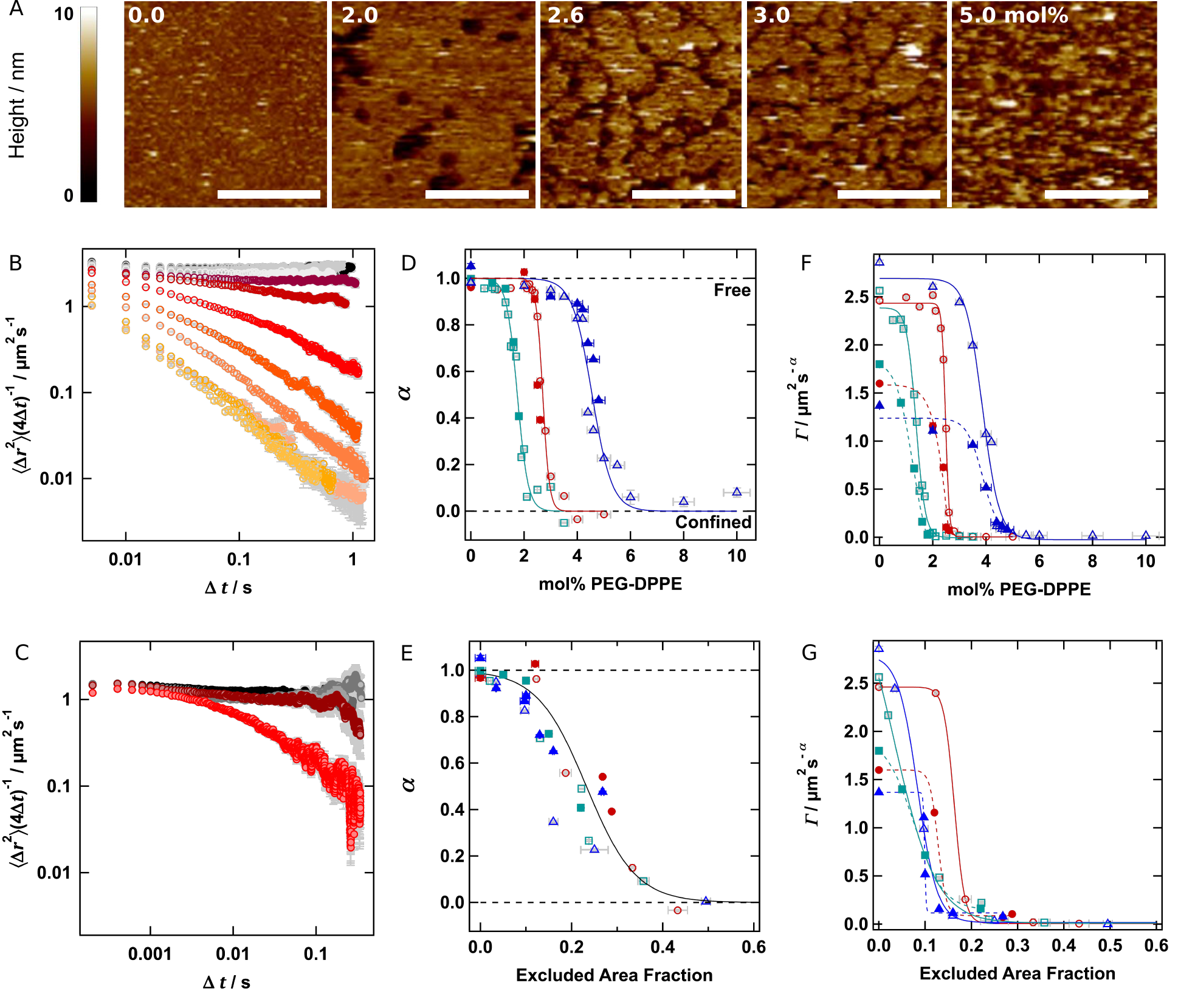}
\caption{
\textbf{Anomalous diffusion in PEG bilayers.} (\textbf{A}) AFM shows an increase in defect area fraction with increasing mol\% PEG-DPPE (scale 500 nm). (\textbf{B}) Anomalous sub-diffusion increases as amount of PEG-DPPE increases from 0 (black) to 10 (yellow) mol\%, here for PEG(2K)-DPPE. (\textbf{C}) Equivalent iSCAT data for 0 to 2.6 mol\% PEG(2K)-DPPE. Variation of $\alpha$ and $\Gamma$ with mol\% PEG-DPPE (\textbf{D\&F}), and excluded area fractions (\textbf{E\&G}).  PEG(1K)- (blue triangles), PEG(2K)- (red circles) and PEG(5K)-DPPE (green squares). Error bars (grey) throughout represent standard errors.} 
\end{figure*}

\section{Results}
\subsection{Supported Lipid Bilayers}
 We produced SUVs from DOPC doped with PEG-DPPE (0 - 10 mol\% PEG-DPPE; 1,2, \& 5 kDa PEG-). Fusion of these SUVs onto a glass coverslip created a SLB. We confirmed the physical nature of the bilayers using AFM (Fig. 2A): As PEG-DPPE content increases, small defects appear in the bilayer. Further increase in the concentration of PEG-DPPE results in the extension of the interfacial defects, until the system crosses the percolation threshold. This leads to confined bilayer patches. Image binarization and autocorrelation were used to calculate the excluded area fraction and the length scale associated with the defects.

The diffusive properties of these bilayers were assessed using single-particle tracking: smTIRF microscopy was used to follow Texas Red-labelled lipids ($\sim 10^{-6}$ mol\% TR-DHPE) at 200 Hz; iSCAT tracked 40 nm antibiotin-conjugated gold nanoparticles (AuNPs) tethered to biotinylated lipids at 5 kHz.

As the concentration of PEG(2K)-DPPE was increased from 0 to 6 mol\%, the gradient of the $\log \left(\langle\Delta r^2 \rangle/4\Delta t\right)$ \emph{vs.} $\log( \Delta t)$ plot deviates from zero (Fig. 2B).  Figure 2C shows the equivalent PEG(2K)-DPPE dataset arising from iSCAT. Similar plots were produced for PEG(1K)- and PEG(5K)-DPPE (Fig. S1). Gradients extracted from these plots ($\alpha-1$) allow calculation of the anomalous exponent, while the y-axis intercept reports the transport coefficient. The values of $\alpha$ for all three PEG molecular weights are collated in Figure 2D; $\alpha$ transitions from 1 (free diffusion) to 0 (confined diffusion). These data were fit empirically by a simple sigmoid. The midpoint of each sigmoid is a measure of the transition between continuous and discontinuous diffusion. Both smTIRF and iSCAT measurements give rise to the same trend (Table S1). Figure 2F shows the equivalent variation of $\Gamma$ with mol\% PEG-DPPE. Again, in agreement with Figure 2D, the apparent diffusion coefficient slows as the particles become confined. It is worth emphasizing that as $\Gamma$ scales with $\alpha$, only points with the same $\alpha$ values can be compared directly. Limiting values for $\Gamma$ differ as expected between smTIRF and iSCAT experiments due to the size difference of  fluorescently labelled lipids and AuNPs \cite{Mascalchi2012}.
 
 Using our AFM calibration (Fig. S2), we are able to convert mol\% to excluded area fraction (Fig. 2E). For both $\alpha$ and  $\Gamma$, the sigmoids for the three different PEG molecular weights now overlap, showing the same trend with excluded area fraction. For $\alpha$ \emph{vs.} excluded area fraction a single sigmoid fit yields a midpoint at $\alpha = 0.232 \pm 0.001$.
 
 \subsection{Monte Carlo Simulations}

\begin{figure*}
	\includegraphics[width=0.85\linewidth]{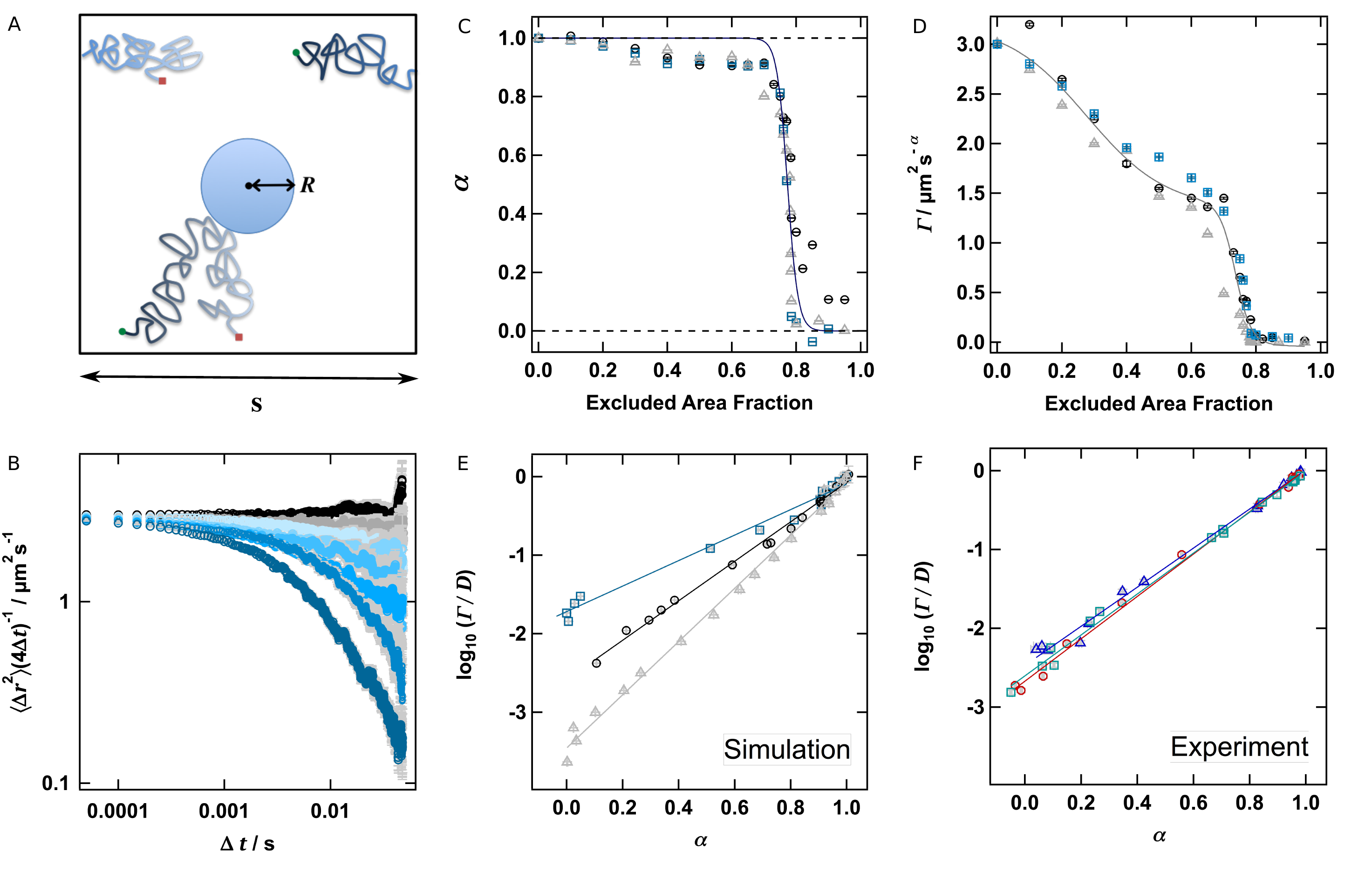}
\caption{\textbf{Monte Carlo simulations of anomalous diffusion}. (\textbf{A}) Schematic of the unit cell. (\textbf{B}) Diffusion analysis of the resultant tracks showed similar behavior to experiment ($R = 500$ nm, $D = 3\; \upmu$m$^2$ s$^{-1}$). (\textbf{C\&D}) A similar trend to experiment was also present for $\alpha$ and $\Gamma$ for $R =$ 150 nm, 500 nm and 1 $\upmu$m (grey triangles, black circles, teal squares respectively). (\textbf{E}) Plots of $\log_{10}\left({\Gamma / D}\right)$ \emph{vs.} $\alpha$ show the expected linear relation, dependent on obstacle size. (\textbf{F}) Similar plots for our experimental data show an essentially static linear relationship for different PEG lengths. PEG(1K)- (blue triangles), PEG(2K)- (red circles) and PEG(5K)- (green squares).}
\end{figure*}

To help improve our understanding of these experimental results we constructed a simple Monte-Carlo simulation of anomalous diffusion: A periodic square lattice of circular, immobile obstacles of radius $R$ was simulated using a unit cell with side-length $s$ (Fig. 4A). A discrete-time random walk was subject to the constraint that the walk cannot enter the circular obstacle. As expected, anomalous behavior arises in the simulation as the excluded area fraction was increased (Fig. 4B), and again a plateau of normal diffusion at short times was observed. When values are collated a sigmoidal trend was present with  $\alpha= 0$ being reached near to the percolation threshold for circular obstacles on a square lattice (0.785).  $\Gamma$ was best fit by a double-sigmoid (Fig. 4D, Table S2).
 
 Combining equations \ref{eqn2} and \ref{eqn3}, a linear variation of $\log_{10}\left({\Gamma / D}\right)$ with $\alpha$ is expected; and reproduced by our data. The characteristic length scales ($\lambda$) calculated from both simulation (Fig. 4E) and experiment (Fig. 4F) are summarized in Table S3.

\section{Discussion}
The presence of PEG-DPPE disrupts SLB formation leaving a network of defects whose area fraction is dependent on the concentration. We have exploited this defect formation to create predictable and tuneable anomalous behavior. It is bilayer continuity (not the presence of PEG as an obstacle) that causes the anomalous behavior. Our controls (Fig. S4) confirm that normal diffusive behavior can be rescued by filling in bilayer defects. 

The variation of $\alpha$ and $\Gamma$ with excluded area fraction that we observe shows a sigmoidal transition between free and confined diffusion. This relationship can be used to tune anomalous behavior. It has been shown using simulations \cite{Stachura2014} and in cell membranes \cite{Murase2004, Schwille1999, Feder1996, Smith1999, Schutz1997} that $\alpha$ values of 0.5 to 0.7 are most biologically relevant. Using our model, we can make specific and controlled changes to obstacle extent that match this range; providing an opportunity to use this simple model to help predict and study biological systems that exhibit complex membrane diffusion \emph{in vivo}.

We must also address the limitations of this model system.  Confinement in cell membranes is not created directly by membrane defects, but is likely due to the excluded area created by membrane proteins and their interactions with lipids. Despite these fundamental differences, both result in a similar restriction to free diffusion in the bilayer, and a parallel can be drawn between the excluded area controlled in this simple model and that inaccessible to diffusing species in cell membranes.

Theory predicts that in a system with finite hierarchy, the diffusion will return to normal behavior (with a reduced diffusion coefficient) at sufficiently long observation times \cite{Saxton2007}. Figure 2B shows that over the time scales observed in these experiments, the diffusion here remains anomalous. As normal behavior returns at around 100 ms for similarly sized compartments in cells \cite{Murase2004},  our model must not possess the restricted range of compartment size that are presumably present in cell membranes. However, we predict that additional control of bilayer defect formation would enable a return to normal diffusion at these timescales, for example by nano-patterning of the substrate before SLB formation \cite{Tsai2008}.

By using two single-particle microscopy techniques we have sampled the anomalous behavior of this model over four orders of magnitude of time. Alone, fluorescence microscopy cannot access the divergence of the diffusivity at short times and the onset of anomalous behavior. However iSCAT is not without its own limitations: iSCAT image analysis requires efficient background subtraction \cite{Kukura2009,Ortega-Arroyo2012a} which fails if particles do not move sufficiently e.g. within the confined regime. The high frame rate of iSCAT also presents its own challenges in data management, preventing us from probing all the relevant timescales with a single technique.

Our simulations helped us to understand the relationship between $\alpha$ and excluded area fraction. In the square lattice model, an excluded area fraction of 0.785 represents the point at which the obstacle diameter is equal to the size of the unit cell, $s$.  This is therefore the point at which confinement occurs. The mid point of the sigmoidal fit occurs at an excluded area fraction of 0.773 and the fit effectively reaches $\alpha$ = 0 at 0.808. If we consider $\alpha$ to sample the probability of being confined, we can say that the mid-point is indicative of the percolation threshold. Using this to interpret the experimental data we find that the percolation threshold of our model falls around an excluded area fraction of $0.232 \pm 0.054$.

Perhaps of greatest interest is the difference between plots of $\log_{10}(D/\Gamma)$ \emph{vs.} $\alpha$ between our simulations and experiment (Table S3): Only our simulation shows a variation in length scale with obstacle size. In contrast, our experimental data shows a similar gradient for all three molecular weights of PEG-DPPE. This suggests that in the experiment the size (but not number or extent) of defects produced are of a similar scale (~150 nm) and are independent of the PEG molecular weight. The values extracted from AFM FWHMs are smaller by around a factor of 3. There is also a modest negative correlation between FWHM and excluded area fraction (see Fig. S4). The differences between experiment and simulation are most likely due to the different topologies for defects in the two cases: To preserve simplicity, defects in the simulation were simple circles. This is in contrast to our experimental AFM images (Fig. 2A) that show not circular, but rather interfacial defects for intermediate PEG-lipid concentrations. We therefore interpret this characteristic length scale as a descriptor of the barrier to diffusion. In our model, that is the scale associated with bilayer defects, and in cells it is likely to be associated with the cytoskeleton. 

\section{Conclusion}
By controlling SLB formation using PEGylated lipids we are able to produce bilayers with defined anomalous diffusive properties, dependent on the excluded area fraction. Thus, we hope our work in part answers the call for a simple and reproducible experimental model, readily tuneable in anomaleity over the length scales observed \emph{in vivo}. This study also opens the way to further experiments that exclude membrane area using more complex methods than the simple inclusion of PEG-lipids presented here. Future work must be directed to expand our understanding of cell membranes - to recreate biological pathways controlled by diffusion, and to enable the rational design of devices with tailored bilayer properties.

\section{Acknowledgements}
We thank the European Research Council for providing funding for this work (ERC-2012-StG-106913 CoSMiC).

\section{References}
\bibliographystyle{apsrev4-1}
\bibliography{Coker_2017}

\newpage

\end{document}



\title{Controlling Anomalous Diffusion in Lipid Membranes\ Supplementary Information}

\author{H. L. E. Coker*}
\author{M. R. Cheetham*}
\affiliation{*Contributed equally to this work.}
\affiliation{Department of Chemistry, Britannia House, King's College London, London SE1 1DB}
\affiliation{Chemistry Research Laboratory, 12 Mansfield Road, University of Oxford, Oxford OX1 3TA}
\author{D. R. Kattnig}
\affiliation{Living Systems Institute \& Department of Physics, University of Exeter, Stocker Road, Exeter EX4 4QD}
\author{Y. J. Wang}
\author{S. Garcia-Manyes}
\affiliation{Department of Physics, Strand Building, King's College London, London WC2R 2LS}
\author{M. I. Wallace}
\affiliation{Department of Chemistry, Britannia House, King's College London, London SE1 1DB}
\date{\today}

\maketitle


\renewcommand{\thefigure}{S\arabic{figure}}
\renewcommand{\thetable}{S\arabic{table}}
\setcounter{figure}{0}
\setcounter{table}{0}

\section{Supplementary Figures}
\begin{figure*}[!htb]
\includegraphics[width=0.8\linewidth]{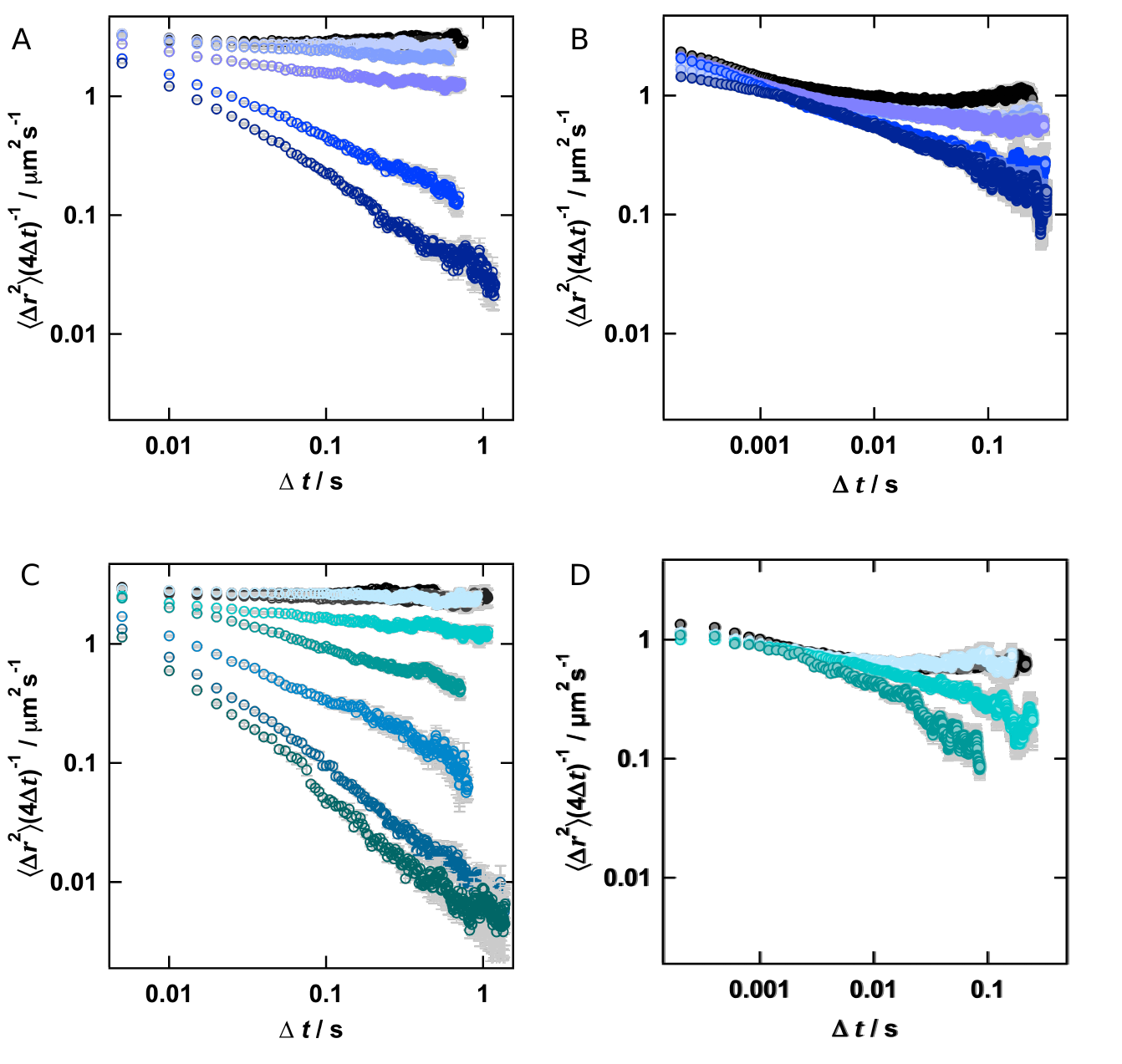}
\caption{{$\left<\Delta r^2\right>/4\Delta t$ \emph{vs.} $\Delta t$ Plots}. PEG(1K)-DPPE; (\textbf{A}) fluorescence and (\textbf{B}) iSCAT and PEG(5K)-DPPE; (\textbf{C}) fluorescence and (\textbf{D}) iSCAT to accompany the PEG(2K)-DPPE data in Fig. 2B/C. The same behaviour is seen for all three PEG molecular weights. Along with Fig. 2B/C the data from these plots are used to extract $\Gamma)$ and $\alpha$ as outlined in the Results section of the main text.}
\label{kusumiplots}
\vspace{5em}
\end{figure*}

\begin{figure*}[]
\includegraphics[width=0.45\linewidth]{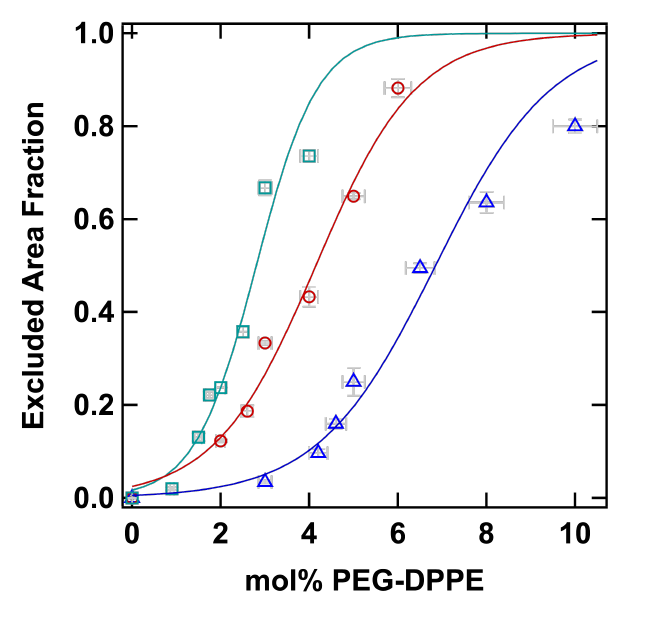}
\caption{ \textbf{Conversion to excluded area fraction}. Image auto-thresholding (see Supplementary Methods) was used to get an excluded area fraction from the AFM images for several concentrations of PEG-DPPE. Plotted against mol\% PEG-DPPE and fitted with sigmoidal functions, we get a calibration curve by which to covert mol\% PEG-DPPE to excluded area fraction.}
\label{calibration}
\vspace{5em}
\end{figure*}

\begin{figure*}[]
\includegraphics[width=0.9\linewidth]{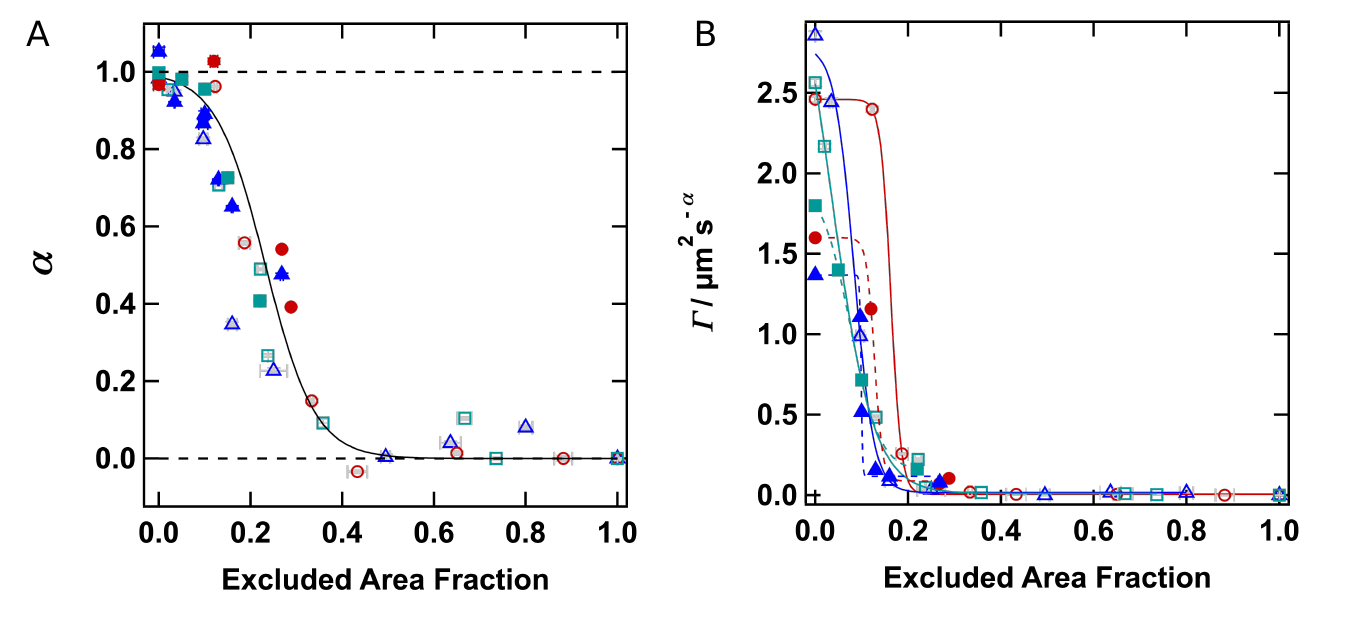}
\caption{ \textbf{Extended plot of (\textbf{A}) excluded area fraction \emph{vs.} $\alpha$ and (\textbf{B}) excluded area fraction \emph{vs.} $\Gamma$}. Fig. 2E \& 2G show the same data for $0\leqslant\alpha\leqslant 0.6$ to highlight the overlap of sigmoids for the three datasets. Here, all data is shown ($0\leqslant\alpha\leqslant 1$). PEG(1K)-, PEG(2K)- and PEG(5K)-DPPE are represented in  blue triangles, red circles and green squares respectively.}
\label{extendedaxis}
\vspace{5em}
\end{figure*}

\begin{figure}[!htb]
\includegraphics[width=1.0\linewidth]{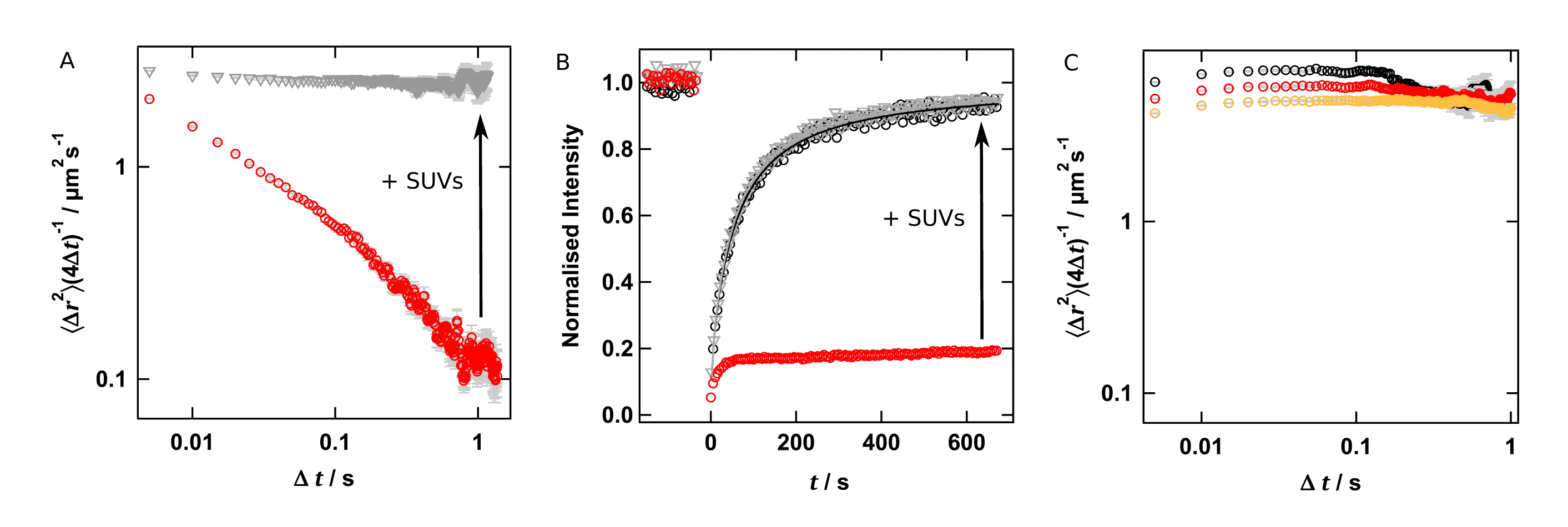}
\caption{\textbf{Anomalous behaviour is a result of confinement in bilayer patches.} (\textbf{A}) single-particle tracking of anomalous diffusion was repeated on  a SLB exhibiting anomalous behaviour (DOPC; 2.6\% PEG(2K)-DPPE).  This bilayer was then incubated with DOPC SUVs. The additional vesicles ruptured within the defects, repaired the bilayer and returned normal diffusive behaviour. (\textbf{B}) FRAP shows similar behaviour, with 2.6\% PEG(2K)-DPPE bilayers exhibiting a large slow and immobile fraction. Upon addition of DOPC SUVs, diffusion again recovers ($D = 1.75\; \upmu$m$^2$ s$^{-1}$, mobile fraction 0.99) to values comparable to a pure DOPC bilayer ($D = 1.81\; \upmu$m$^2$ s$^{-1}$, mobile fraction 1.00).(\textbf{C}) Droplet interface bilayers (DIBs) \cite{Bayley2008, Leptihn2013} (Fig. S5) were used to create unsupported lipid bilayers and confirm PEG-lipids alone do not result in anomalous behaviour. In DIBs, defects would result in conductance across the bilayer, unstable droplets and ultimately bilayer rupture \cite{Sengel2016}. Following our previous methods \cite{Leptihn2013}, `lipid-in' DIBs (2.6 or 5.0 mol\% PEG(2K)-DPPE; $\sim 10^{-6}$ mol\% TR-DHPE) were formed. Single-particle tracking using TIRF microscopy showed normal diffusion even for 5.0 mol\% PEG-DPPE, well above the threshold value for anomalous behaviour in SLBs. Bilayers containing 0 (black), 2.6 (red) and 5.0 (yellow) mol\% PEG(2K)-DPPE all exhibit normal diffusion.
In Droplet Interface Bilayers (DIBs), }	
\vspace{5em}
\end{figure}

\begin{figure*}[]
\includegraphics[width=0.45\linewidth]{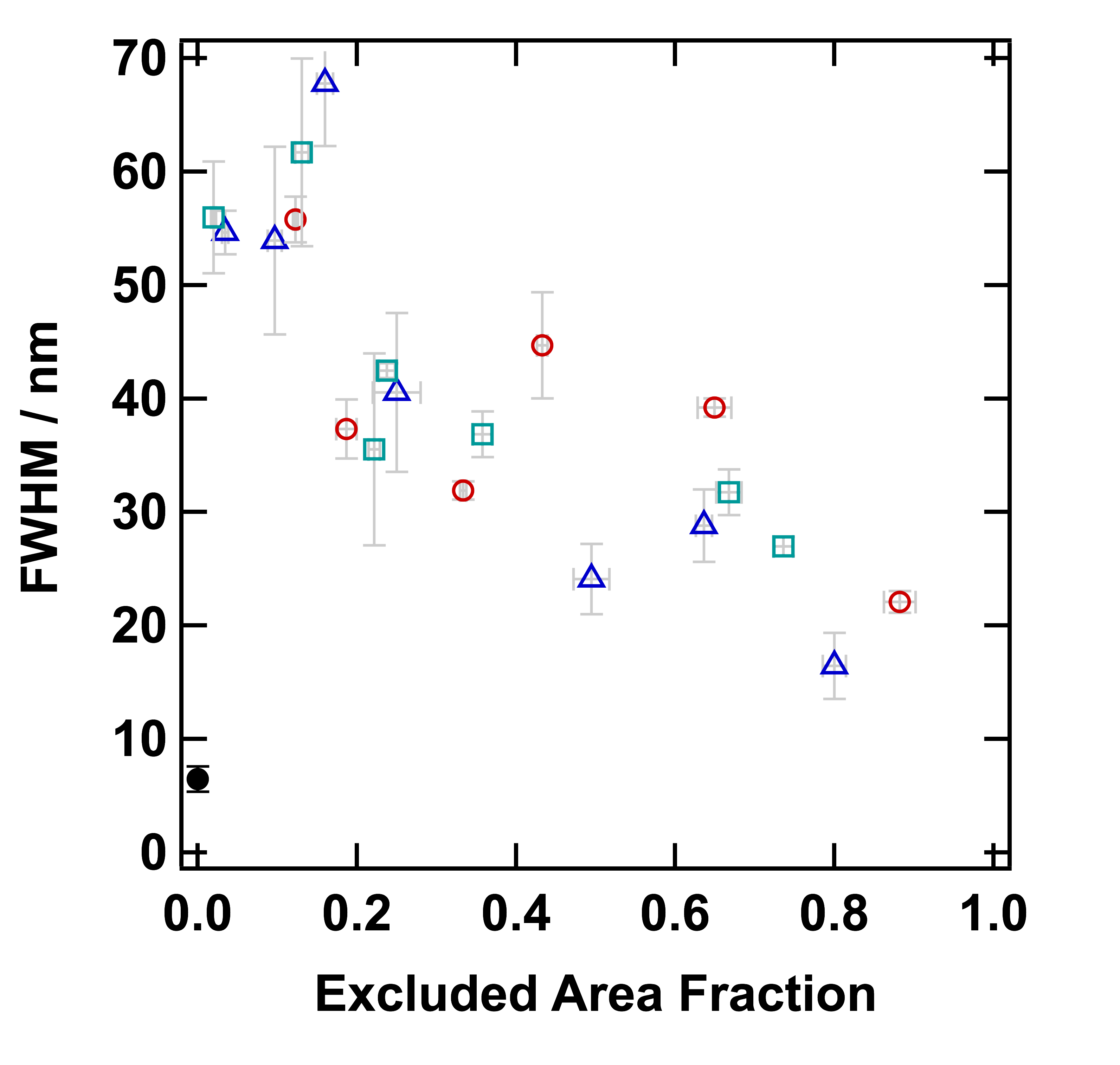}
\caption{ \textbf{Length scales extracted from autocorrelation analysis of the AFM images}. There is a modest decrease in length scale as excluded areas fraction increases. Black data point is DOPC only SLB. PEG(1K)-, PEG(2K)- and PEG(5K)-DPPE are represented in  blue triangles, red circles and green squares respectively.}
\label{AFM}
\vspace{5em}
\end{figure*}

\begin{figure*}
\includegraphics[width=0.45\linewidth]{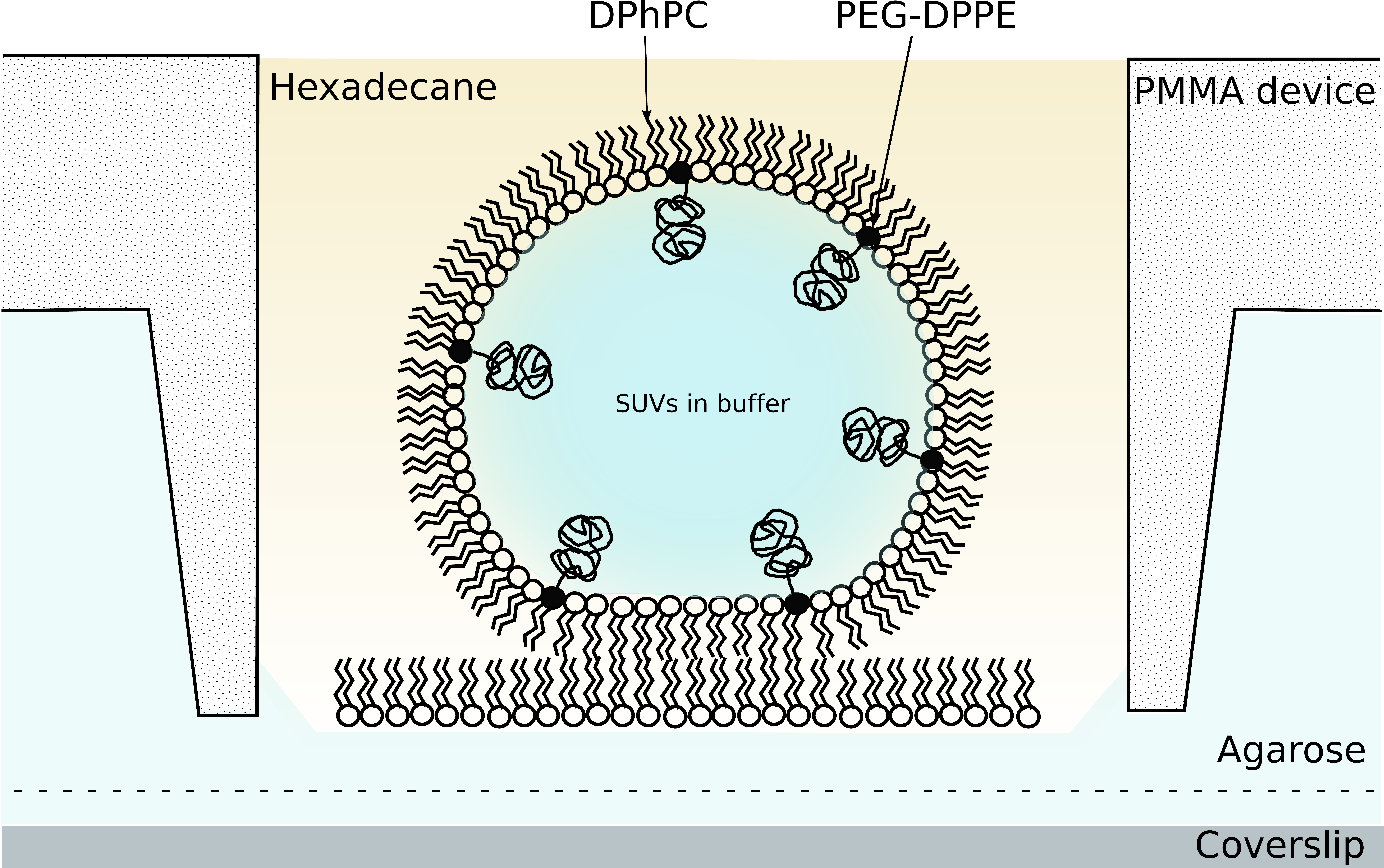}
\caption{ \textbf{Schematic of a droplet interface bilayer (DIB).} DIBs were prepared following our previous protocol (Leptihn \emph{et al.} Nature Protocols \textbf{8}, 1048, 2013).}
\label{DIB}
\vspace{5em}
\end{figure*}

\clearpage

\section{Supplementary Tables}
\begin{table}[!htb]
\begin{ruledtabular}
\caption{\textbf{Experiment sigmoid fit midpoints.} Midpoints extracted from the sigmoid fit of the $\alpha$ \emph{vs.} mol\% PEG-DPPE plots, Fig. 2D, for both fluorescence and iSCAT data. In each case the midpoint agrees within 6\% and can be regarded as an indicator of the percolation threshold.}
\label{expmid}
\begin{tabular}{llll}
 PEG / kDa & mol\% (fluorescence) & mol\% (iSCAT) \\
1 & $4.556 \pm 0.004$ & $4.785 \pm 0.053$   \\
2 & $2.700 \pm 0.002$ & $2.545 \pm 0.019$   \\
5 & $1.745 \pm 0.001$ & $1.745 \pm 0.005$   \\
\end{tabular}
	\end{ruledtabular}
     \vspace{1em}
\end{table}

\begin{table}[!htb]
\begin{ruledtabular}
\caption{\textbf{Simulation sigmoid fit midpoints.} Midpoints extracted from the sigmoid fit of the $\alpha$ \emph{vs.} EAF (Fig. 4C) and $\Gamma$ \emph{vs.} EAF (Fig. 4D) plots. Figure 4C was fitted with a single sigmoid, whereas Figure 4D was better fit by a double sigmoid.}
\label{simmid}
\begin{tabular}{llll}
  & EAF 1 &  EAF 2 \\
$\alpha$ (Fig. 4C) & - & $0.773 \pm 0.002$  \\
$\Gamma /\ \mu\textrm{m}^2 \textrm{ s}^{-\alpha}$ (Fig. 4D) & $0.277 \pm 0.001$ & $0.737 \pm 0.001$   \\
 \end{tabular}
     \end{ruledtabular}
     \vspace{1em}
\end{table}

\begin{table}[!htb]
\begin{ruledtabular}
 \caption{\textbf{Anomalous length scales ($\lambda$) for experiment and simulation.}}
\label{lengths}
  \begin{tabular}{lllll}
  \textbf{Simulation}\\
 R /nm & 150 & 500 & 1000    \\
 $\lambda$ /nm & $65 \pm 3$ & $177 \pm 7$ & $479 \pm 24$   \\
 \\
 \textbf{Experiment}\\
PEG /kDa & 1 & 2 & 5   \\
$\lambda$ /nm & $196 \pm 4$ & $145 \pm 2$ & $160 \pm 3$    \\
AFM FWHM /nm & $54 \pm 8$ & $56 \pm 2$ & $62 \pm 8$
 \end{tabular}
     \end{ruledtabular}
     \vspace{5em}
\end{table}

\clearpage
\newpage
\section{Supplementary Methods}
\subsection{Materials}
1,2-dioleoyl-\emph{sn}-glycero-3-phosphocholine (DOPC), 1,2-diphytanoyl-\emph{sn}-glycero-3-phosphocholine (DPhPC) and 1,2-dioleoyl-\emph{sn}-glycero-3-phosphoethanolamine-N-cap-biotinyl (biotin-DOPE) were purchased from Avanti Polar Lipids (Alabaster, AL). Texas Red 1,2-dihexadecanoyl-\emph{sn}-glycero-3-phosphoethanolamine (TR-DHPE) triethylammonium salt and 1,2-dipalmitoyl-\emph{sn}-glycero-3-phosphoethanolamine-N-[methoxy(polyethylene glycol) - 1000/2000/5000] ammonium salt (PEG1/2/5K-DPPE) were purchased from Invitrogen (Eugene, OR). Goat antibiotin-conjugated 20 nm gold nanoparticles were purchased from BBI (Cardiff, UK). Standard phosphate-buffered saline tablets (P4417), hexadecane (296317) and silicone oil AR20 (10836) were purchased from Sigma-Aldrich (St-Louis, MO). All chemicals and solvents were of analytical grade and used without further purification. 18.2 M$\Omega$ cm water (Merck Millipore, Billerica, MA) was used to prepare all solutions. 
 
 \subsection{Supported Lipid Bilayers}
Glass coverslips (VWR, Menzel Gl{\"a}ser \#1) were rigorously cleaned using stepwise sonication with detergent (Decon-90), water, and propan-2-ol. Immediately before use, the glass was dried under nitrogen, cleaned with oxygen plasma for 3 minutes (Diener Electronic, Femto), and stored under water before use. A well was created on each coverslip using vacuum grease (Dow Corning high vacuum grease). 50 $\upmu$L of SUV stock was diluted 1:1 in buffer (PBS pH 7.4) and added to the chamber immediately. The vesicles were incubated for 60 minutes before the membranes were washed with de-gassed Milli-Q. 

SLBs were prepared on glass coverslips by vesicle fusion \cite{Brian1984}. Lipid mixtures  (1 mg mL$^{-1}$ DOPC;  $\sim 10^{-6}$ mol\% each TR-DHPE and biotin-DOPE) were doped with varying amounts of PEG-lipids to span a range of 0 to 10 mol\%. Lipid mixtures were dried under nitrogen and placed under vacuum overnight. The dried lipids were hydrated in water and votexed before tip sonication for 15 minutes. The resulting clear, vesicle suspension was centrifuged  (3 minutes; 14000\emph{g}) before the supernatant was separated and retained. 

 \subsection{Droplet Interface Bilayers}
 DIBs were prepared following our previous protocol \cite{Leptihn2013}. Briefly, SUVs of DPhPC only and DPhPC + PEG(2K)-DPPE were prepared as described above and diluted with buffer to a final concentration of 0.5 mg ml$^{-1}$. 100 nl droplets of SUV solution were pipetted into a microfabricated tank containing hexadecane and incubated for 40 min.  DPhPC (8.9 mg ml$^{-1}$ in 9:1 hexadecane:silicone oil) was prepared from chloroform stock. A microfabricated PMMA device containing the agarose layer on a glass coverslip was incubated with the DPhPC in oil solution for 20 min. Incubated droplets were then added to the device and left for 10 min to allow bilayer formation before imaging. The DIB experiment is represented schematically in Figure S6.
 
 \subsection{Atomic Force Microscopy}
 AFM was carried out using a Bioscope (Bruker, MA). An SNL-10B cantilever was used with a spring constant of 0.12 N m$^{-1}$. The instrument was run in peak-force tapping mode with a maximum applied force of 40 pN. Auto-thresholding was carried out on exported TIFs in FIJI \cite{Schindelin2012}. Auto-correlation of the images was performed using the FD Math function. The resultant plot was then radially averaged using a radial profile. The resultant plot was fitted with a Lorentzian function to extract the FWHM.
 
 \subsection{Fluorescence Recovery After Photobleaching}
 Fluorescence recovery after photobleaching (FRAP) experiments were carried out as described by Axelrod \textit{et al.} \cite{Axelrod1976}. Briefly, an DOPC SLB was made with 0.5 mol\% TR-DHPE, and subjected to intense laser light through a small iris for ~5 s. The iris was then opened, the illumination intensity reduced, and the bilayer was imaged every 5 s for the following 20 min. The result was a video showing the fluorescence slowly recovering in the region where the intense laser light had caused strong photobleaching. Image analysis of this recovery process yielded the diffusivity of the TR-DHPE in the SLB. This was quantified by fitting with a modified Bessel function as outlined by Soumpasis \cite{Soumpasis1983}. This fitting method is only applicable to bilayers exhibiting normal diffusion and as such, the bilayer containing PEG(2K)-DPPE at 2.6 mol\% was not quantified.

 \subsection{Total Internal Reflection Fluorescence Microscopy}
A 532 nm continuous-wave laser was launched into an inverted microscope (Eclipse TiE; Nikon) and  focussed at the back aperture of the objective lens (60$\times$— TIRF oil-immersion NA 1.49, Nikon, $\sim$1.4 kW cm$^{-2}$ at the sample) such that total internal reflection occurred. The resultant fluorescence signal was transmitted through dichroic (ZT532rdc) and bandpass (605/55) filters (both Chroma, Bellows Falls, VT) before being imaged onto an electron-multiplying CCD (iXon+ 860; Andor) at 200 Hz.

\subsection{Interferometric Scattering Microscopy}
 A custom-built microscope was built to conduct these measurements as described \cite{Kukura2009,OrtegaArroyo2016}. Briefly, a 639 nm laser beam (Toptica, Graefelfing, Munich) was directed through a polarizing beam splitter  and quarter wave plate (both Thorlabs, Newton, NJ) before reaching a focus at the back aperture of an objective lens (100$\times$ oil immersion NA 1.49 Nikon). Light scattered from the  coverslip interface and the object of interest, produce an interference pattern, which is returned through the beam splitter. The resultant signal was then magnified (overall magnification 174$\times$) and imaged onto a CMOS camera (Phantom Miro 340). To track AuNP-labelled lipids,  a SLB was formed as described above and 5 $\upmu$L of anti-biotin OD10 40 nm gold nanoparticles were added. After 2 hours incubation,  the SLB was washed thoroughly with buffer to remove unbound nanoparticles before imaging.  Image stacks of 10000 frames were recorded at 100 kHz. Image stacks were then temporally-averaged to 5 kHz before tracking tracking.

\subsection{Single-Particle Tracking}
The movement of fluorescent lipids between frames of the recorded videos was tracked using the TrackMate plugin for ImageJ \cite{Tinevez2016}. Briefly, spots were detected in each frame using a Laplacian of Gaussian (LoG) filter. Tracks were then generated by linking these detected spots together using a simple linear assignment problem (LAP) tracker. The output from this was a collection of tracks containing the space-time co-ordinates of each point in the track. This data was subsequently used to obtain mean-squared displacements for different observation times, which in turn was used to obtain diffusivity values via the random walk model of diffusion.

\subsection{Monte Carlo Simulations}
For each of 200 simulated particles, two separate random walks were created (corresponding to the displacement in $x$ and $y$ coordinates) using a pseudorandom number generator (Mersenne Twister). In order to account for the presence of obstacles to diffusion, periodic boundary conditions were used. The unit cell of the simulation is shown in Figure 4A of the main text. With each step of the random walk, the new coordinates were tested to see if they were inside an obstacle or not. If they were, then that step was rejected, and a new one generated. The resulting sets of co-ordinates for each simulated particle were subsequently processed in the same way as experimental data obtained from tracking labelled particles. These tracks represent diffusing particles that have been hindered due to the presence of obstacles, and exhibited anomalous diffusion as expected.

\bibliographystyle{apsrev4-1}

\bibliography{Coker_2017}